# HTCondor data movement at 100 Gbps

Igor Sfiligoi
University of California
San Diego
La Jolla, CA, USA
isfiligoi@sdsc.edu

Frank Würthwein
University of California
San Diego
La Jolla, CA, USA
fkw@ucsd.edu

Thomas DeFanti
University of California
San Diego
La Jolla, CA, USA
tdefanti@ucsd.edu

John Graham
University of California
San Diego
La Jolla, CA, USA
jjgraham@ucsd.edu

*Abstract*—HTCondor is a major workload management system used in distributed high throughput computing (dHTC) environments, e.g., the Open Science Grid. One of the distinguishing features of HTCondor is the native support for data movement, allowing it to operate without a shared filesystem. Coupling data handling and compute scheduling is both convenient for users and allows for significant infrastructure flexibility but does introduce some limitations. The default HTCondor data transfer mechanism routes both the input and output data through the submission node, making it a potential bottleneck. In this document we show that by using a node equipped with a 100 Gbps network interface (NIC) HTCondor can serve data at up to 90 Gbps, which is sufficient for most current use cases, as it would saturate the border network links of most research universities at the time of writing.

## I. INTRODUCTION

Scientific computing needs are continually growing in time, with many problems becoming intractable on single nodes and requiring a distributed computing approach. A major problem of distributed computing is data movement, as data and compute resources are not co-located anymore. While shared filesystems can hide the problem from users, they are not a panacea; data is still being moved around and scaling them across wide area networks is notoriously hard.

HTCondor [1], a popular distributed high throughput (dHTC) workload management system used by the Open Science Grid (OSG) [2], foregoes the need for a shared filesystem, providing native data movement for the managed compute jobs. This capability allows HTCondor to aggregate compute resources with minimal constraints, requiring neither storage mounting privileges nor advanced network privileges. Indeed, most OSG compute resources come from nodes that allow only UNIX-like processes and are behind restrictive firewalls, e.g., no incoming networking allowed.

In a default HTCondor setup, data flows in and out of the submit node, which also holds the compute job queue and where users have login privileges. For maximum performance, the storage should be local to such a node, although other solutions can be used for either cost or resiliency reasons. Since all data transfers associated with the managed compute jobs flow through such a node, it can become a bottleneck, especially with very spiky workload patterns. We thus measured the capabilities of HTCondor as data movement tool on state-of-the-art network hardware, which at the time of writing was represented by a 100 Gbps network interface (NIC).

This work was partially funded by the US National Research Foundation (NSF) under grants OAC-2030508, OAC-1836650, OAC-1826967 and OAC-1541349.

Pre-print version, July 2021

Benchmarking was performed on the Pacific Research Platform's (PRP) Nautilus [3] environment, which is briefly described in section 2. We measured sustained 90 Gbps network traffic on local area networks and 60 Gbps across the US, with detailed benchmarking results presented in section 2 and 3.

## II. THE TEST ENVIRONMENT

HTCondor is very scalable, typically serving tens of thousands of worker nodes managing compute resources. Nevertheless, the data transfers only happen at job boundaries, so for data transfer benchmarking purposes, only the job startup rate is important, not the total pool size. For the purpose of this paper, we assumed that there were approximately 200 slots that need file transfer at any point in time, which is what one would expect in a pool with 20k slots serving jobs lasting 6 hours, each spending 3 minutes in file transfer. We simulated that by running jobs with trivial runtimes but large input data.

The test hardware was accessed by means of the PRP, a Kubernetes-based system spanning the US (and beyond), with all nodes connected with high-speed network links. The most demanding, i.e. HTCondor submit node with a 100 Gbps NIC was located at the University of California San Diego campus (UCSD), while the worker nodes executing the jobs could be located anywhere. We ran several tests, including a test with all nodes inside UCSD and a test with all nodes on the US east coast, whose results are described in the next two sections.

The PRP made performing these tests trivial; no special privileges were needed to deploy HTCondor services spanning the US. The HTCondor workers are just regular containers launched as unprivileged pods in Kubernetes. The other HTCondor services could also be launched as unprivileged pods, but we ran into a performance issue following that path. Since PRP uses Calico to establish the virtual private network (VPN) in Kubernetes, which is required for unprivileged use, we noticed that the HTCondor submit node was bottlenecked by the VPN overhead, limiting the throughput to about 25 Gbps. The majority of the benchmarking tests were thus run using a HTCondor submit container running without VPN, which does require additional privileges, but allowed us to exceed 90 Gpbs. We will investigate more user-friendly workarounds in future work.

## III. BENCHMARKING ON LOCAL AREA NETWORK

Since all tests were executed on a shared network setup, we first measured the HTCondor data movement performance inside a local area network, where we were expecting only minor network interference from other activities.

For this first test, all nodes were located inside UCSD. To minimize the number of nodes involved, all nodes were equipped with a 100 Gbps NIC, both the one running the HTCondor submit pod and the six HTCondor worker nodes. The worker nodes were configured to provide a grand total of 200 execute slots. We used the latest available stable HTCondor version, which at the time of writing was version 9.0.1. We also used the default security settings, which resulted in all file transfers being fully authenticated, AES encrypted, and integrity checked.

On the submit node, we created a single 2GB file with random content and then created 10k unique file names hard linking to it. From the user point of view, we thus had 10k independent files, while from the storage point of view there was a single 2GB data area, that easily fit into the system cache. The purpose of the exercise was to measure the HTCondor data movement performance and this setup guaranteed that the storage subsystem was not the bottleneck. The compute power was provided by an 8-core AMD EPYC 7252 CPU.

The main test consisted in submitting 10k jobs as a single HTCondor submit transaction, each pointing to a unique input file and a short-running validation script. We collected both the HTCondor logs and network monitoring data, which showed that we were using on about 11 GBps, or 90 Gbps of network bandwidth, as seen in the screenshot available in Fig. 1.

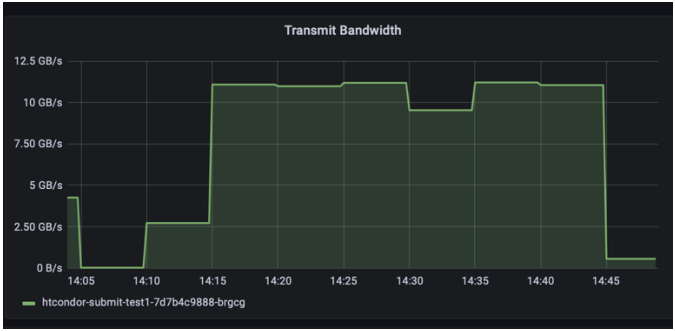

Fig. 1. Screenshot of the PRP network monitoring web page during the local area HTCondor test. Each bin represents the average over 5 minutes.

All jobs finished within 32 minutes. The median job runtime was 5 seconds, and the median input data transfer time was 2.6 minutes. Output file transfer times were negligible, and no errors were encountered.

The above experiment shows that HTCondor is capable of almost saturating a 100 Gbps NIC if the storage subsystems can feed it fast enough. Note that for the above test we disabled the HTCondor file transfer queuing mechanism, which is by default tuned for much slower spinning disk storage systems. Using the default settings, a similar 10k job test completed in 64 minutes, i.e. in about double the time.

## IV. BENCHMARKING ON THE WIDE AREA NETWORK

Established that HTCondor can almost saturate a 100 Gbps NIC on a local area network, we moved to measure its performance over the wide area network (WAN). We kept the submit node at UCSD, which is located in California state, and deployed pods on nodes that were as far away as possible while still having at least a 100 Gbps WAN network path to UCSD.

All used worker nodes were located in New York state. Only one node had a 100 Gbps NIC, while another four had a 10 Gbps NIC. The round trip time between the submit and worker nodes was about 58 ms, and traversed network links operated by CENIC, Internet2 and NYSERNet.

Apart from using different hardware resources, the test setup was virtually identical to the one described in the previous section. In this test HTCondor managed to use about 7.5 GBps, or 60 Gbps of network bandwidth, as seen in the screenshot available in Fig. 2. All jobs finished in 49 minutes and the median input data transfer time was 3.3 minutes; all other metrics were comparable.

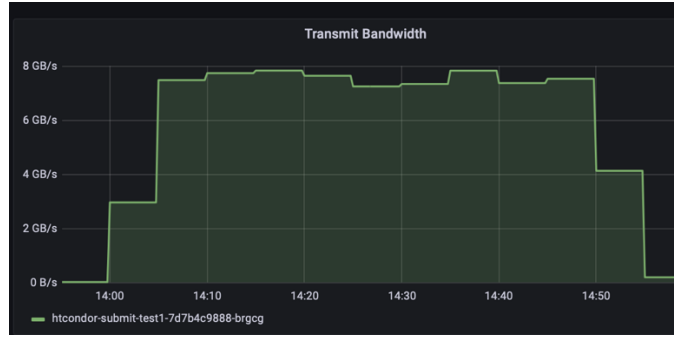

Fig. 2. Screenshot of the PRP network monitoring web page during the wide area HTCondor test. Each bin represents the average over 5 minutes.

Given the shared nature of the wide area networking, we are not disappointed by the lower throughput. Unfortunately we however cannot completely rule out HTCondor being the main bottleneck, as we do not have monitoring information for all of the network switches along the path.

## V. SUMMARY

We show that HTCondor architecture of explicitly managing data movement is not a bottleneck in current use cases, as it can scale the data throughput to tens of Gbps, if paired with a 100 Gbps NIC and sufficiently performant storage subsystem. In our tests, executed with HTCondor version 9.0.1 and on hardware managed by the PRP, we measured sustained data throughputs of 90 Gbps on LAN and 60 Gbps on cross-US network links. All with end-to-end strong authentication, encryption, and integrity checks, available to all users without any additional setup or configuration steps.

ACKNOWLEDGMENT

This work was partially funded by the US National Research Foundation (NSF) under grants OAC-2030508, OAC-1836650, OAC-1826967 and OAC-1541349.